\documentclass[iop]{emulateapj_jw}

\usepackage{hyperref}
\usepackage{epsfig}
\usepackage{natbib}
\usepackage{graphicx}                
\usepackage{verbatim}              
\usepackage{amsmath,amsthm,amsfonts,amsopn,amssymb} 
\usepackage{pdflscape}
\usepackage{afterpage}
\usepackage{rotating}					
\usepackage{ulem}
\usepackage{epstopdf}  
\usepackage{multirow}
\usepackage{color}
\usepackage[left]{lineno}
\received{}
\accepted{}

\slugcomment{to appear in AJ}
\shorttitle{The Rich Get Richer}
\shortauthors{Wang}

\begin{document}

\title{Revealing A Universal Planet-Metallicity Correlation For Planets of Different Sizes Around Solar-Type Stars}
\author{
Ji Wang\altaffilmark{1} and
Debra A. Fischer\altaffilmark{1}
} 
\email{ji.wang@yale.edu}
%
\altaffiltext{1}{Department of Astronomy, Yale University, New Haven, CT 06511 USA}

\begin{abstract}

The metallicity of exoplanet systems serves as a critical diagnostic of planet formation mechanisms. Previous studies have demonstrated the planet-metallicity correlation for large planets ($R_P\ \geq\ 4\ R_E$); however, a correlation has not been found for smaller planets. With a sample of 406 $Kepler$ Objects of Interest whose stellar properties are determined spectroscopically, we reveal a universal planet-metallicity correlation: not only gas-giant planets ($3.9\ R_E\ < R_P\ \leq\ 22.0\ R_E$) but also gas-dwarf ($1.7\ R_E\ < R_P\ \leq\ 3.9\ R_E$) and terrestrial planets ($R_P\ \leq\ 1.7\ R_E$) occur more frequently in metal-rich stars. 
The planet occurrence rates of gas-giant planets, gas-dwarf planets, and terrestrial planets are $9.30^{+5.62}_{-3.04}$, $2.03^{+0.29}_{-0.26}$, and $1.72^{+0.19}_{-0.17}$ times higher for metal-rich stars than for metal-poor stars, respectively.

\end{abstract}


\section{Introduction}

Since thousands of exoplanets and exoplanet candidates have been discovered~\citep{Schneider2011,Wright2011}, we are now in a position to study the statistical attributes of the ensemble of known exoplanets. The correlations between the exoplanets' occurrence and the properties of their host stars help us to understand how planets form and evolve. For example, studies found that gas-giant planets occur more frequently around metal-rich stars~\citep{Gonzalez1997, Santos2004, Fischer2005, Johnson2010}. This planet-metallicity correlation provides a critical diagnostic for planet formation mechanisms: the correlation for gas-giant planets is consistent with the core accretion scenario for exoplanets that orbit closer than 20 AU around main sequence stars~\citep{Ida2004, Mordasini2009, Alibert2011}. Although recent studies confirmed a planet-metallicity correlation around sub-giant stars, a planet-metallicity correlation was not found for giant stars~\citep{Takeda2008,Ghezzi2010,Maldonado2013}. Furthermore, the planet-metallicity correlation was not found for planets smaller than gas-giant planets, such as Neptune-like planets and rocky planets~\citep{Sousa2008, Bouchy2009, Mayor2011,Neves2013}. 

In addition to the planet-metallicity correlation, correlations between the planet occurrence rate and other stellar properties were suggested.~\citet{Johnson2010} found a positive correlation of the gas-giant planet occurrence rate with stellar mass, but the correlation did not emerge in the analysis in~\citet{Mortier2013}.~\citet{Eggenberger2004, Eggenberger2011} found that circumstellar gas-giant planets are less frequent in systems with close-in stellar binaries. Subsequently studies further confirmed the influence of stellar multiplicity on planet formation~\citep[e.g.,][ and references therein]{Wang2014b,Wang2014}. These correlations help us to build a complete physical picture of planet formation under different conditions. 

The NASA $Kepler$ mission~\citep{Borucki2010} has been a tremendous success in finding transiting exoplanets. More than 7000 planet candidates have been discovered during its 4.5-year mission\footnote{http://exoplanetarchive.ipac.caltech.edu}~\citep{Borucki2011,Batalha2013,Burke2013}. Although the planet occurrence rate has been extensively studied~\citep{Catanzarite2011, Youdin2011, Traub2012, Howard2012, Fressin2013, Dressing2013}, the dependence of the planet occurrence rate on metallicity is a relatively less-visited topic for the $Kepler$ data. 

~\citet{Schlaufman2011} found that gas-giant planets are preferentially found around metal-rich stars. While they did not find a planet-metallicity correlation for small-radius planets around solar-type stars, they reported a positive correlation between stellar metallicity and the occurrence rate of small-radius planet candidates around late K dwarfs. Their finding was later interpreted as a result of mis-classifications of giant stars to dwarf stars~\citep{Mann2012, Mann2013}. ~\citet{Buchhave2012} measured metallicity for a sample of 152 $Kepler$ planet host stars and confirmed the planet-metallicity correlation for gas-giant planets. They found that planets with radii smaller than 4 $R_E$ have a wide range of metallicity with median close to the solar value, and concluded that there was no planet-metallicity correlation for planets smaller than 4 $R_E$.~\citet{Everett2013} took spectra of 220 faint $Kepler$ planet host stars ($K_P > 14$) and reached a similar conclusion to that of~\citet{Buchhave2012}. Recently, ~\citet{Buchhave2014} expanded their metallicity measurements to 406 $Kepler$ planet host stars, and found that metallicity is important in regulating planet structure. In their recent data, the average metallicities for gas giant planets and gas dwarf planets are above the solar metallicity ($0.18\pm0.02$ dex and $0.05\pm0.01$ dex), the average metallicity for terrestrial planets is consistent with the solar metallicity at $-0.02\pm0.02$ dex. These intriguing results motivate us to revisit the planet-metallicity correlation using the newly available spectroscopically-determined stellar and planetary properties. 

The planet-metallicity correlation for small planets has not been confirmed and quantified: we still do not know whether and how the small planet occurrence rate changes as a function of stellar metallicity. The challenge is both the number of planets and the number of stars being searched. The former used to be small; the latter is often difficult to extract. With hundreds of spectroscopically-characterized $Kepler$ planet candidates and hundreds of thousands of $Kepler$ stars, we now have a unique sample of planet host stars and stars with non-detections. This sample enables us to further investigate the planet-metallicity correlation for planets of different sizes. We describe our sample, methodologies in \S \ref{sec:Method}. Result and comparison with previous works are discussed in \S \ref{sec:comp}. Summary and discussion are given in \S \ref{sec:Summary}.

\section{Sample and Method}
\label{sec:Method}

\subsection{Sample}
\label{sec:sample}

Our method relies on knowing stellar properties for both planet host stars and stars with no planet detections. For the planet host stars, we used stars in~\citet{Buchhave2014} whose stellar properties are determined via the Stellar Parameter Classification (SPC) technique~\citep{Buchhave2012}. Typical precision for SPC is 50K, 0.10 dex, and 0.08 dex for $\rm{T}_{\rm{eff}}$, log(g), and [Fe/H], respectively. We selected the stars in~\citet{Buchhave2014} hosting planets with orbital periods less than 100 days, which account for 93\% (376/406) of their sample. 

Since a large number ($>180,000$) of stars have been observed by the $Kepler$ mission~\footnote{The Barbara A. Mikulski Archive for Space Telescopes, http://archive.stsci.edu/kepler/}, only a tiny fraction of $Kepler$ stars received spectroscopic followup observations, the majority of $Kepler$ stars have only photometrically-determined stellar properties~\citep{Brown2011,Huber2014}. Therefore, for stars with no planet detections, we adopted the stellar parameters from the NASA Exoplanet Archive\footnote{http://exoplanetarchive.ipac.caltech.edu} and converted the values into the more representative stellar properties to remove systematic errors (see \S \ref{sec:KicSpec} for details). We restricted our sample to be solar-type dwarf stars ($4800\ K<\rm{T}_{\rm{eff}}\leq6500$ $K$, log(g) $\ge 4.2$). 

\subsection{KIC vs. SPC}
\label{sec:KicSpec}
\begin{figure}
\begin{center}
\includegraphics[width=16cm,height=5.0cm]{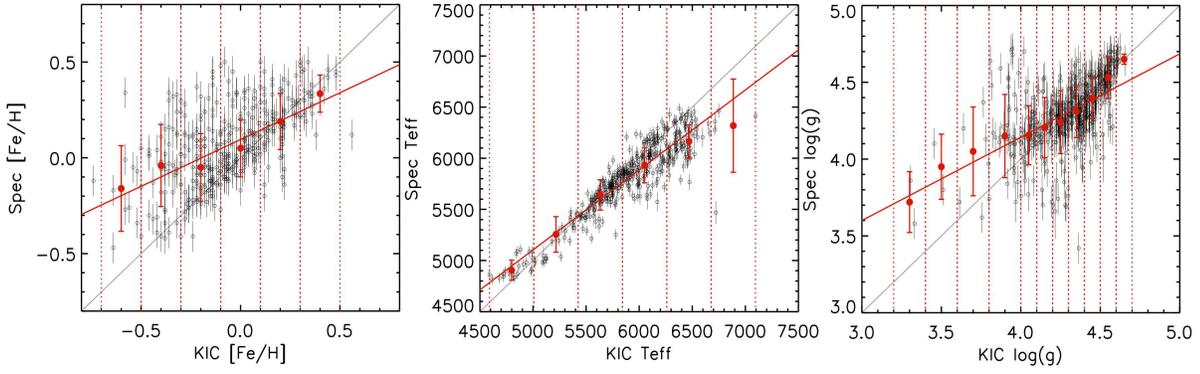} \caption{Comparison of stellar properties from the KIC and the SPC analysis (left: metallicity, middle: effective temperature, right: surface gravity). Red circles and error bars are medians and standard deviations for data points in bins of stellar properties. Each bin is defined by two adjacent red dotted lines. Solid lines indicate 1:1 ratio, and red lines are the best linear fits. 
\label{fig:SME_KIC}}
\end{center}
\end{figure}

Stellar properties for most $Kepler$ stars are from the $Kepler$ Input Catalog~\citep[KIC,][]{Brown2011}. The KIC values are known to have large uncertainties and systematic errors. To investigate the uncertainty and the systematic error, we compared the KIC values with the spectroscopically-determined values for the 406 stars in~\citet{Buchhave2014} (shown in Fig. \ref{fig:SME_KIC}). While we found a large dispersion ($\sim0.2$ dex) of spectroscopically-determined metallicities at a given KIC metallicity, there is a positive correlation between the SPC and KIC metallicities, [Fe/H]$_{\rm{SPC}}$=0.10+0.49$\times$[Fe/H]$_{\rm{KIC}}$. The positive correlation suggests that, although not accurate, the KIC metallicity can be used as a proxy of the stellar metallicity.~\citet{Dawson2012} used the KIC metallicity to find evidence of planet-planet interactions for giant-planet migration in metal-rich stars. Similarly, for $\rm{T}_{\rm{eff}}$ and log(g), we also found a positive correlation between SPC and KIC values, $\rm{T}_{\rm{eff}, SPC}$=1194+0.78$\times$$\rm{T}_{\rm{eff, KIC}}$, and log(g)$_{\rm{SPC}}$=1.97+0.54$\times$log(g)$_{\rm{SPC}}$, respectively. The correlations for [Fe/H], $\rm{T}_{\rm{eff}}$ and log(g) can be used to remove systematic errors, and convert KIC values to values that better represent stellar properties. 

We describe how we converted KIC values to the more representative values as follows. Using the metallicity as an example, we divided KIC metallicity into 6 bins with width of 0.2 dex starting at -0.7 dex, and calculated the median value and the standard deviation of the SPC metallicities in each bin (shown as red filled circles and error bars in Fig. \ref{fig:SME_KIC}). A more representative metallicity than the KIC metallicity can be calculated by interpolating between adjacent red points. Similarly, a more representative error bar can also be estimated. For $\rm{T}_{\rm{eff}}$, we used 6 bins with width of  417.5 K starting at 4590 K to cover the entire $\rm{T}_{\rm{eff}}$ range. For log(g), we used 10 bins with width of 0.1 or 0.2 dex starting at 3.2 dex. The width is changed from 0.1 dex to 0.2 dex for bins with log(g) lower than 4.0 dex because of small number of available data points. We will use the more representative stellar properties for $Kepler$ stars with no planet detections throughout the paper. 

\subsection{Dividing Sample on the $R_P$-[Fe/H] Plane}
\label{sec:KicSpec}
\begin{figure}
\begin{center}
\includegraphics[width=16cm,height=10cm]{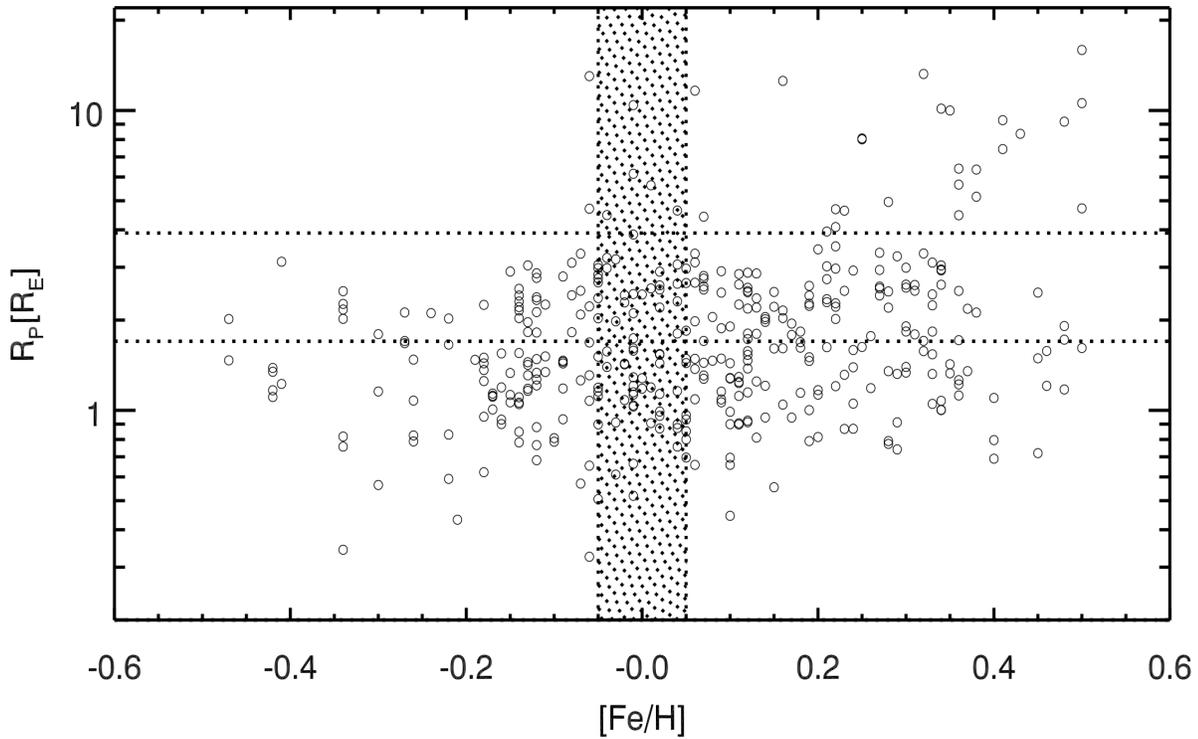} \caption{$Kepler$ planet candidates on the [Fe/H]-$R_P$ plane. The plane is divided into 6 sub-regions based on metallicity and planet radius. The dotted area is the metallicity buffer zone ($-0.05\leq$ [Fe/H] $\leq0.05$). Stars in the buffer zone are excluded in our analysis. 
\label{fig:Rp_FeH}}
\end{center}
\end{figure}

We defined two metallicity groups, metal-poor group with [Fe/H] $<-0.05$ and metal-rich group with [Fe/H] $>0.05$. The metallicity range between -0.05 and 0.05 serves as a buffer zone, limiting the contamination between two metallicity groups. The width of the buffer zone is consistent with the typical error bar quoted in~\citet{Buchhave2014}. 
There are $\sim53,000$ and $\sim49,000$ stars in the metal-poor and metal-rich groups, respectively. Given the uncertainty of the converted metallicity ($\sim0.2$ dex), we did not further divide the sample along the metallicity dimension. 
In each metallicity group, we further divided the stars with detected planets into three sub-groups based on planet size: gas-giant planets ($3.9\ R_E\ < R_P\ \leq\ 22.0\ R_E$), gas-dwarf planets ($1.7\ R_E\ < R_P\ \leq\ 3.9\ R_E$), and terrestrial planets ($R_P\ \leq\ 1.7\ R_E$). The scatter plot of planet radius and stellar metallicity is shown in Fig. \ref{fig:Rp_FeH}. 

\subsection{Fraction of Stars With Detected Planets}
\label{sec:face_value}

We adopted the following Monte Carlo method to estimate the fraction of stars with detected planets. For each $Kepler$ star with no planet detection, we assigned the star with stellar properties ($\rm{T}_{\rm{eff}}$, log(g), and [Fe/H]) based on the converted, more representative values. The values are then perturbed assuming Gaussian distributions with standard deviations that are equal to the more representative error bars as shown in Fig. \ref{fig:SME_KIC}. For each $Kepler$ planet host star in our sample, we perturbed the measurements of stellar properties ($\rm{T}_{\rm{eff}}$, log(g), and [Fe/H]) and planet properties ($R_P$) with the reported measurement uncertainties. With this set of simulated data, we counted the number of stars with detected planet candidates ($N_P$) and the number of searched stars ($N_S$) in each sub-region as shown in Fig. \ref{fig:Rp_FeH}. We then calculated the ratio of $N_P$ and $N_S$, a ratio which we refer to as the fraction of stars with detected planets. We repeated this procedure 1000 times and calculated the median values and the standard deviations for $N_P$ and $N_S$. Since $N_P$ and $N_S$ were counted in each iteration, we considered counting errors which were assumed to follow Poisson distribution. Therefore, the final uncertainties of  $N_P$ and $N_S$ are the summation in quadrature of the standard deviation and the Poisson noise. This approach takes into account both the statistical uncertainty and the uncertainties of stellar and planetary properties.

The fractions of stars with detected planets are given in Table \ref{tab:planet_metallicity}. The reported values and uncertainties are the mode and the 68\% credible interval of the $N_P/N_S$ distribution. We emphasize that measurements of the planet occurrence rates for planets of different sizes require further information such as detection sensitivity and survey incompleteness. Thus, the numbers that we list should not be construed to be the estimates of the true planet occurrence rates. However, the ratio of $N_P/N_S$ between two metallicity groups is a measure of the ratio of the planet occurrence rates of two metallicity groups, i.e., the relative planet occurrence rate of metal-rich to metal-poor stars. This is true only if  the detection sensitivity and survey incompleteness affect the two metallicity groups in the same way.   

\subsection{Bias Against Planet Detection For Metal-Rich Stars}
\label{sec:bias}
\begin{figure}
\begin{center}
\includegraphics[width=16cm,height=16.5cm]{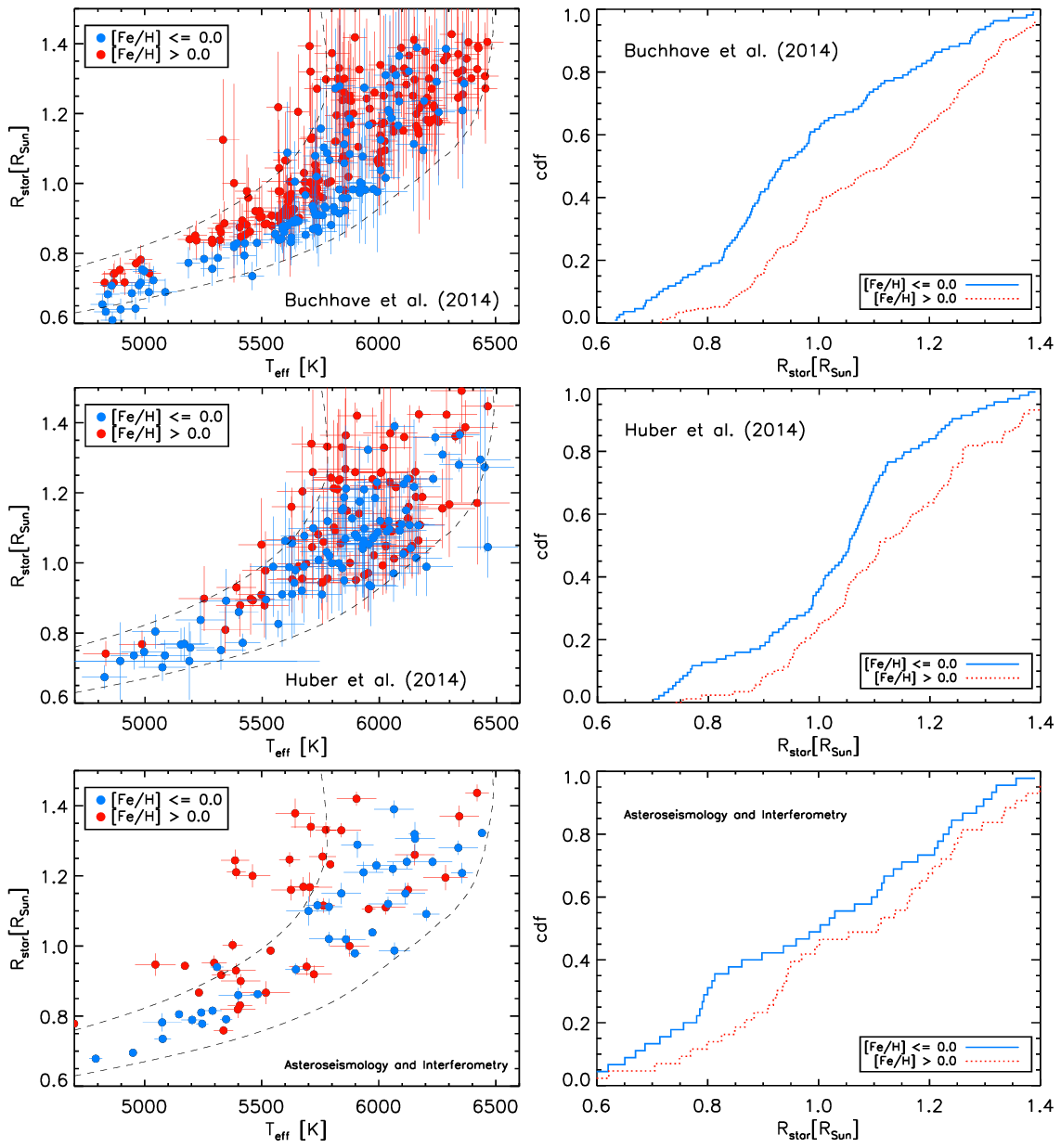} \caption{Left panels: stellar radii vs. T$_{\rm{eff}}$ for KOIs from ~\citet{Buchhave2014} (top), $Kepler$ stars with asteroseismic and spectroscopic measurements from ~\citet{Huber2014} but excluding stars with the SPC analysis (middle), and stars with asteroseismic and interferometric measurements from~\citet{Chaplin2014},~\citet{Boyajian2012},~\citet{Boyajian2013}, and~\citet{vonBraun2014} (bottom). Dashed lines in each sub-plot on left panels are predictions of stellar radii from stellar evolution model~\citep{Dotter2008} at the age of 5 Gyr for two metallicities, [Fe/H]=-0.4 (lower) and [Fe/H]=0.4 (upper). Right panels are comparisons of radius distributions between metal-rich and metal-poor stars. Metal-rich stars are on average $\sim$5-20\% larger than metal-poor stars at a given T$_{\rm{eff}}$.
\label{fig:Rs_FeH}}
\end{center}
\end{figure}

At a given photometric precision, it is more difficult to detect a transiting planet around a larger star, so stellar radius affects the planet detection sensitivity. To investigate whether detection sensitivity is the same for stars from two metallicity groups, we checked the dependence of stellar radius on metallicity. Top panels in Fig. \ref{fig:Rs_FeH} show stars with spectroscopic followup observations from~\citet{Buchhave2014}. We limited the comparison for stars with $4800\ K<\rm{T}_{\rm{eff}}\leq6500$ $K$ and log(g) $\ge 4.2$, which is consistent with our sample selection criteria. We found that metal-rich stars are $\sim$20\% larger than metal-poor stars. We performed two-sided K-S test, the p value is $1.8\times10^{-5}$, and the maximum difference of cumulative distribution is 0.29. Two-sided Kuiper test, which is more sensitive to tails of distributions, shows a similar result with a p value of $7\times10^{-4}$. Next, we checked if the stellar radius dependence on metallicity is a systematic for the SPC method. We collected stars with spectroscopically- or asteroseismically-determined stellar properties from~\citet{Huber2014}, but excluded stars from~\citet{Buchhave2012}. These stars consist of a sample whose stellar properties are determined free of the systematics (if any) of the SPC method. With this sample of 182 stars, we again found that metal-rich stars are larger; the median radius of the metal-rich sample is $\sim$5\% larger than that for the metal-poor sample. The K-S test gives a p value of 0.006. However, the Kuiper test is inconclusive (p = 0.089 $\geq 0.01$). Finally, we checked all stars whose radii are measured with asteroseismic and interferometric methods~\citep[bottom panels in Fig. \ref{fig:Rs_FeH},][]{Boyajian2012,Boyajian2013,vonBraun2014,Chaplin2014}. In this case, we found that metal-rich stars are $\sim$5\% larger than metal-poor stars, but neither the K-S test (p=0.14) nor the Kuiper test (p=0.41) is significant. However, the maximum difference between two cumulative distributions is 0.24, similar to two previous cases. The smaller sample size (N=88) may explain the inconclusive tests.


In a transiting planet survey like the $Kepler$ mission, planet detection is more difficult around larger stars for a given planet size, so there is a bias against planet detection for metal-rich stars whose radii are larger than metal-poor stars. This bias may not be an issue for large planets, whose signals can be detected regardless of the 5\%-20\% stellar radius difference. However, the detection bias gets stronger for terrestrial planets and small gas-dwarf planets; the transiting signals of these planets may be only marginally detected given the $Kepler$ precision. Therefore, the ratio of the fraction of stars with planets for metal-rich stars to that for metal-poor stars reflects a lower limit of the relative planet occurrence rate of metal-rich stars to metal-poor stars.

\section{Planet Occurrence Rate vs. Stellar Metallicity}
\label{sec:comp}

\subsection{Result}
\label{sec:res}

\begin{figure}
\begin{center}
\includegraphics[width=16cm,height=12cm]{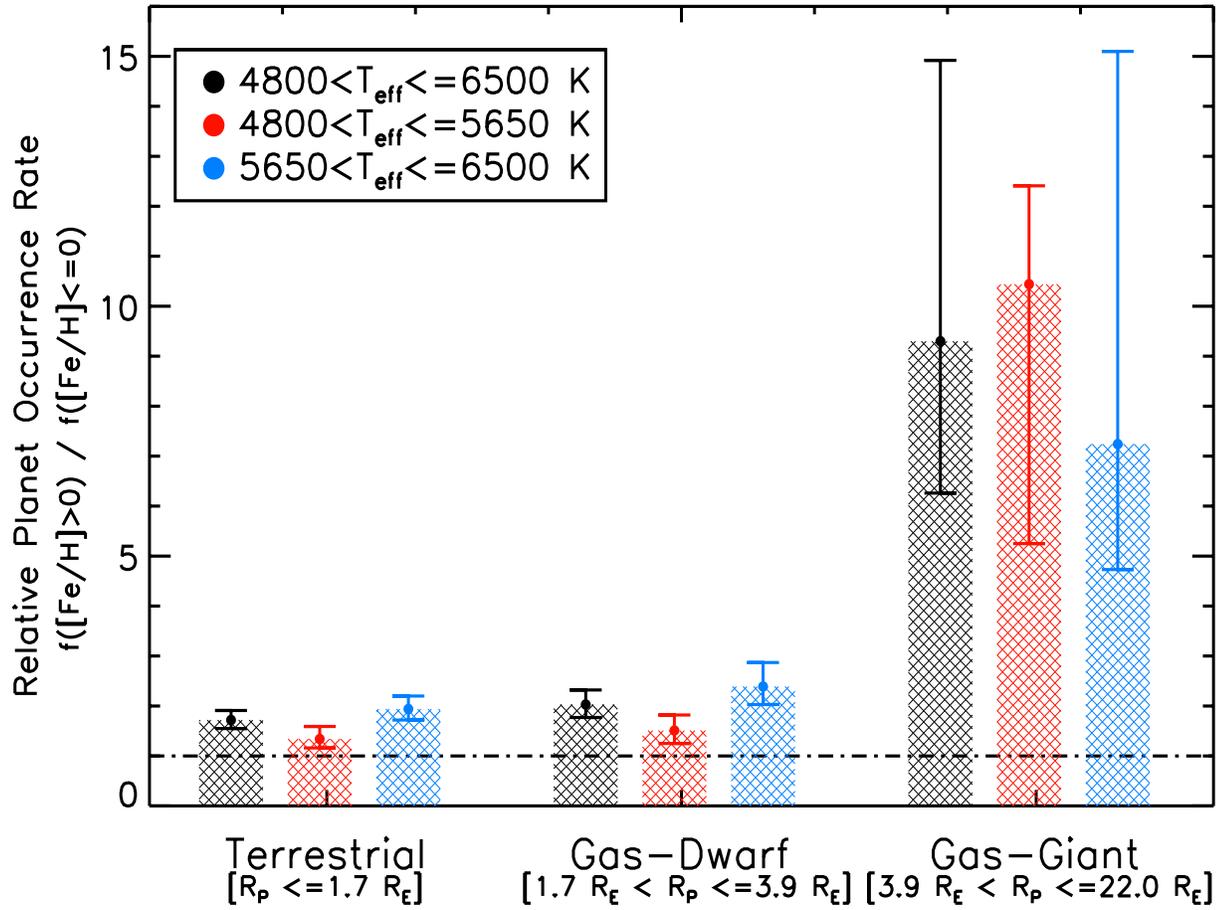} 
\caption{The relative planet occurrence rate as a function of planet size. The relative planet occurrence rate is the ratio of the planet occurrence rate for metal-rich stars to metal-poor stars. A value of 1 (dash-dotted line) indicates no metallicity dependence of the planet occurrence rate. The relative planet occurrence rates for gas-giant planets, gas-dwarf planets, and terrestrial planets are all significantly higher than 1, indicating the planet-metallicity correlation is universal for planet of all sizes. For terrestrial and gas-dwarf planets, the relative planet occurrence rate for stars with higher effective temperatures (blue) is higher than stars with lower effective temperatures (red). 
\label{fig:Fraction_Rp}}
\end{center}
\end{figure}

We found a universal planet-metallicity correlation: not only gas-giant planets but also gas-dwarf planets and terrestrial planets occur more frequently in metal-rich stars (Fig. \ref{fig:Fraction_Rp}). The dependence of the planet occurrence rate on metallicity decreases with decreasing planet size. The planet occurrence rates of gas-giant planets, gas-dwarf planets, and terrestrial planets are $9.30^{+5.62}_{-3.04}$, $2.03^{+0.29}_{-0.26}$, and $1.72^{+0.19}_{-0.17}$ times higher for metal-rich stars than for metal-poor stars, respectively. Given the detection bias against metal-rich stars (\S \ref{sec:bias}), these values are the lower limits for the planet-metallicity correlation: the dependence of the planet occurrence rate on metallicity may be stronger.

We further divided the sample into two with different effective temperature ranges, one with $4800\ K<\rm{T}_{\rm{eff}}\leq5650$ $K$, and the other one with $5650\ K<\rm{T}_{\rm{eff}}\leq6500$ $K$. The dividing $\rm{T}_{\rm{eff}}$ was selected to match the mean $\rm{T}_{\rm{eff}}$ of the sample. For the sub-sample with lower $\rm{T}_{\rm{eff}}$, we found that the relative planet occurrence rates of metal-rich to metal-poor stars are $10.44^{+1.97}_{-5.19}$, $1.51^{+0.31}_{-0.26}$, and $1.34^{+0.25}_{-0.18}$ for gas-giant planets, gas-dwarf planets, and terrestrial planets, respectively. The relative planet occurrence rates in the sub-sample with higher $\rm{T}_{\rm{eff}}$ are $7.24^{+7.86}_{-2.51}$, $2.39^{+0.48}_{-0.36}$, and $1.94^{+0.26}_{-0.22}$ for gas-giant planets, gas-dwarf planets, and terrestrial planets, respectively. The relative planet occurrence rates for gas-giant planets are similar between two $\rm{T}_{\rm{eff}}$ sub-samples, but they are different for smaller planets (Fig. \ref{fig:Fraction_Rp}). This may suggest that, for smaller planets, the metallicity dependence of the planet occurrence rate is weaker for stars with lower $\rm{T}_{\rm{eff}}$ than for stars with higher $\rm{T}_{\rm{eff}}$. However, the hypothesis replies heavily on more observations in the future to confirm or refute. 

\subsection{Statistical Test}
\label{sec:test}

To understand the likelihood of the above result being produced by a random sample, we repeated the Monte Carlo simulations in \S \ref{sec:face_value}. Instead of using the metallicities from~\citet{Buchhave2014}, we randomly assigned metallicities to planet host stars following the more representative metallicity distribution of the overall $Kepler$ stellar population. In only 1 out of 1000 trials we observed a value higher than 1.72, which is the relative planet occurrence rate for terrestrial planets. We did not get a higher value than the reported relative planet occurrence rate for gas-dwarf and gas-giant planets. The result indicates that it is highly unlikely that the universal planet-metallicity correlation is mimicked by a random sample. For a sample of planet host stars with randomized metallicities, the relative planet occurrence rates of metal-rich to metal-poor stars are $1.07^{+0.50}_{-0.40}$, $1.06^{+0.20}_{-0.18}$, and $1.06^{+0.21}_{-0.19}$ for gas-giant planets, gas-dwarf planets, and terrestrial planets, respectively. All the values are consistent with unity, which is what we expect from a randomized sample with no planet-metallicity correlation. 


\subsection{Comparison to Previous Results}
\label{sec:comp_pre}

The positive correlation of the gas-giant planet occurrence rate with stellar metallicity is consistent with previous results based on the $Kepler$ data~\citep{Schlaufman2011, Buchhave2012, Everett2013}. The median [Fe/H] for $Kepler$ metal-poor and metal-rich stars in our sample is -0.14 dex and 0.23 dex, respectively. Our result indicates that a change of 0.37 dex in [Fe/H] results in a change of $9.30^{+5.62}_{-3.04}$ times in the gas-giant planet occurrence rate. In comparison, an exponential law dependence with a power $\sim$2.0~\citep{Fischer2005, Udry2007} predicts that the boost of the planet occurrence rate due to metal enhancement is $\sim$5.5. Considering the uncertainties of the metallicity measurement and the exponential power, our result for gas-giant planets is consistent with previous studies. 

For terrestrial planets, we found that the planet occurrence rate for metal-poor stars is lower than that for metal-rich stars at 4.2 $\sigma$ level. In comparison, 
~\citet{Buchhave2012} found that the average metallicity for stars hosting planets smaller than 2 $R_E$ is comparable to the solar value, which is a necessary condition for a null-correlation between plant occurrence and stellar metallicity. However, their finding is not sufficient to conclude the null-correlation without knowing the overall metallicity distribution of $Kepler$ stars. ~\citet{Buchhave2014} measured metallicities for a larger sample of planet host stars. For stars with terrestrial planets ($R_P\ \leq\ 1.7\ R_E$), the median metallicity is 0.04 dex. Among 207 stars with terrestrial planets, 122 have metallicities above the solar value and 85 have metallicities equal or lower than the solar value. As the sample size increases, the median metallicity for stars with terrestrial planets becomes higher than the solar value although it is still consistent with the solar value. 

We investigated whether the metallicity distribution of the overall $Kepler$ stars is statistically identical to that for stars hosting terrestrial planet. Assuming the more representative metallicities for $Kepler$ stars based on the comparison shown in Fig. \ref{fig:SME_KIC}, we compared the metallicity distribution of $Kepler$ stars to that for stars with terrestrial planets. The K-S test (p = 0.005 $\leq$ 0.01) rejects the null hypothesis that the metallicity distribution for terrestrial planet host stars is drawn from the overall $Kepler$ stars. The Kuiper test is inconclusive (p = 0.05 $\geq 0.01$) although hinting the same result as the K-S test. This test shows the importance of taking into consideration the metallicity distributions of both planet host stars and the overall stellar sample. From the difference in metallicity distribution, we went through the analyses described in this paper, and found that terrestrial planets are $1.72^{+0.18}_{-0.17}$ times more abundant around metal-rich stars than metal-poor stars. 

The finding of a positive correlation with metallicity for small planets was also mentioned by ~\citet{Schlaufman2011}, but they found this correlation for late-K dwarfs, which we did not consider in our sample. ~\citet{Schlaufman2011} did not find a significant planet-metallicity correlation for earlier spectral type. However, they used $g-r$ colors as a proxy to stellar metallicity, which are not as precise as spectroscopically-determined values, and is prone to contamination of giant stars~\citep{Mann2012}. 

The positive planet-metallicity correlation for terrestrial and gas-dwarf planets is seemingly at odds with other previous studies from Doppler surveys~\citep{Udry2007, Sousa2008, Bouchy2009, Mayor2011}. However, there are several possible explanations for the disagreement. First, small sample size may cause the insignificant planet-metallicity correlation.~\citet{Mayor2011} studied 23 Doppler planets less massive than 30 M$_\oplus$. They found that the median of their host stars metallicities matches with the median metallicities of the overall stellar sample. To investigate the effect of a small sample size, we downsized our sample to mimic their analysis. We randomly selected 822 $Kepler$ stars to simulate their overall stellar sample, and 23 planet host stars ($R_P\ \leq$ 3.9 $R_E$) to simulate their low-mass planet host star sample, Then, we compared the metallicity distributions of these two samples using the K-S test and the Kuiper test. We found that only in 7.3\% (for the K-S test) and 2.5\% (for the Kuiper test) of trials the p values are smaller than 0.01. Therefore, the non-detection of a planet-metallicity correlation in~\citet{Mayor2011} may be attributed to a small sample size for low-mass planet host stars. 

Second, the weaker dependence of the planet occurrence rate on metallicity for stars with lower effective temperatures may be another possible explanation. If~\citet{Mayor2011} sample is dominated by stars with $\rm{T}_{\rm{eff}}\leq5650$ $K$, then the planet-metallicity correlation is not as significant as it is for stars with higher $\rm{T}_{\rm{eff}}$. The relative planet occurrence rate for these lower $\rm{T}_{\rm{eff}}$ stars is less than 2 $\sigma$ away from unity, which is the value corresponding to no planet-metallicity correlation (see Fig. \ref{fig:Fraction_Rp}). To find the fraction of stars with $\rm{T}_{\rm{eff}}\leq5650$ $K$, we checked the HARPS FKG star sample~\citep{Sousa2011}, 55\% of stars in the sample have $\rm{T}_{\rm{eff}}$ lower than 5650 $K$. In comparison, 37.5\% of the $Kepler$ solar-type stars have $\rm{T}_{\rm{eff}}$ lower than 5650 $K$. The relative portion of stars with $\rm{T}_{\rm{eff}}\leq5650$ $K$ is thus higher for the HARPS sample than for the $Kepler$ sample. We therefore expect a less metallicity dependence of planet occurrence rate for the HARPS sample. We conducted a similar simulation to test small size effect on stars with lower $\rm{T}_{\rm{eff}}$, only 4.9\% (for the K-S test) and 1.8\% (for the Kuiper test) of trials returned p values that indicate significant difference (p $\leq$ 0.01). Therefore, the difference in sample size and $\rm{T}_{\rm{eff}}$ distribution may explain the discrepancy of the planet-metallicity correlation for small planets between the $Kepler$ survey and previous Doppler surveys.

\section{Summary and Discussion}
\label{sec:Summary}

\subsection{Summary}
\label{sec:sum}
With a sample of 406 $Kepler$ planet host stars taken from~\citet{Buchhave2014}, we compared the stellar properties from the KIC and those from the SPC analysis. From the comparison, we derived a method to convert the KIC values for stellar properties to more representative values. Based on the more representative stellar properties, we divided stars into metal-poor and metal-rich groups, and counted the number of stars ($N_S$) and the number of planet host stars ($N_P$) that belong to these two metallicity groups. Next, we calculated the fraction of stars with planets ($N_P/N_S$). We used the ratio of the fractions of stars with planets for metal-rich to metal-poor stars as an estimation of the relative planet occurrence rate, a quantity evaluating the planet occurrence rate change due to metal enhancement. We found a universal planet-metallicity correlation for solar-type stars ($4800\ K<\rm{T}_{\rm{eff}}\leq6500$ $K$, log(g) $\ge 4.2$): planets of all sizes are more abundant around metal-rich stars than they are around metal-poor stars. The relative planet occurrence rates for metal-rich stars to metal-poor stars are  $9.30^{+5.62}_{-3.04}$, $2.03^{+0.29}_{-0.26}$, and $1.72^{+0.18}_{-0.17}$ for gas-giant planets ($3.9\ R_E\ < R_P\ \leq\ 22.0\ R_E$), gas-dwarf planets ($1.7\ R_E\ < R_P\ \leq\ 3.9\ R_E$), and terrestrial planets ($R_P\ \leq\ 1.7\ R_E$), respectively. Since we found that metal-rich stars are 5\%-20\% larger than metal-poor stars, there is a detection bias against planets around metal-rich stars, so the relative planet occurrence rates reported in this paper are the lower limits.

\subsection{The Choice of Buffer Zone}
\label{sec:buffer}

We chose a metallicity buffer zone with width of 0.1 dex, which is comparable to the reported error from~\citet{Buchhave2014}. However, the error for the stars without spectroscopic observations may be higher than 0.1 dex. We changed the width of the buffer zone to investigate how the width changes our results. For a buffer zone width of 0.15 dex, the relative planet occurrence rates for metal-rich stars to metal-poor stars are  $12.13^{+12.46}_{-4.21}$, $2.20^{+0.31}_{-0.29}$, and $1.81^{+0.21}_{-0.19}$ for gas-giant planets, gas-dwarf planets, and terrestrial planets, respectively. For a buffer zone width of 0.20 dex, the relative planet occurrence rates for metal-rich stars to metal-poor stars are  $13.91^{+11.45}_{-5.07}$, $2.41^{+0.43}_{-0.33}$, and $1.92^{+0.22}_{-0.21}$ for gas-giant planets, gas-dwarf planets, and terrestrial planets, respectively. Widening the buffer zone has two effects. First, the widening increases the metallicity difference between the metal-rich and metal-poor sample and thus magnifies the planet-metallicity correlation. Second, the widening decreases the number of available data points and thus increases the uncertainty. As the buffer zone width increases, a higher relative planet occurrence rate is observed for each type of planet, although the numbers overlap within error bars. The increasing relative planet occurrence rate with increasing buffer zone width is another piece of evidence for the universal planet-metallicity correlation. Widening the buffer zone leads to two samples that are more separated by metallicity, which reveals a stronger metallicity dependence of planet occurrence rate, i.e., a higher relative planet occurrence rate. 

\subsection{Disk Time Scale and Small Planet Formation}
\label{sec:disk}

In \S \ref{sec:res}, we reported a weaker dependence of the small planet occurrence rate on metallicity for stars with lower effective temperatures ($4800\ K<\rm{T}_{\rm{eff}}\leq5650$ $K$). The relative planet occurrence rates of metal-rich to metal-poor stars are  $1.51^{+0.31}_{-0.26}$ and $1.34^{+0.25}_{-0.18}$ for gas-dwarf planets and terrestrial planets, respectively. In comparison, the relative planet occurrence rates for stars with higher $\rm{T}_{\rm{eff}}$ ($5650\ K<\rm{T}_{\rm{eff}}\leq6500$ $K$) are $2.39^{+0.48}_{-0.36}$ and $1.94^{+0.26}_{-0.22}$ for gas-dwarf planets and terrestrial planets, respectively. This may be explained by a competition effect between the formation of planetesimals and the dispersal of dust materials in disk. Dust in a protoplanetary disk is dispersed with a time scale of several Myr~\citep{Haisch2001,Mamajek2009}.  Observations of transition disks demonstrate that the central region of the circumstellar disk clears before the outer regions~\citep[e.g.,][]{Isella2009}, and imaging surveys of the brightest disks in the nearby Ophiuchus star forming region demonstrate that central clearings are very common in even the most massive disks at ages of a few Myr~\citep{Andrews2009,Andrews2010}. The observed properties of transitional disks are consistent with clearing due to the dynamical influence of a low-mass companion embedded in the disk~\citep[e.g.,][]{Andrews2011}, although photo-evaporation is likely to play a role in disk dispersal, possibly at later stages~\citep[e.g.,][]{Alexander2013}.  There is also some evidence that the dust disks around intermediate mass stars evolve more rapidly than those around solar-type stars~\citep[e.g.,][]{Yasui2014}.  Therefore, planetesimal formation may be cut off at an earlier time in the disks around more massive stars. In order to rapidly form larger planetesimals that are immune to the effects of photo-evaporation and the reduced accretion flow across the gap formed by low-mass companions embedded in the disk, systems with higher metallicity are favored in forming close-in terrestrial and gas-dwarf planets around more massive stars. By comparison, the clearing process may take place slightly later for less massive stars, such that planetesimals can afford to form more slowly. Therefore, metallicity dependence for close-in terrestrial and gas-dwarf planets around less massive stars is not as strong as it is for more massive stars.

\noindent{\it Acknowledgements} We acknowledge anonymous referees for their insightful comments. We thank Tabetha Boyajian for her stimulating comments and providing interferometric measurements of stellar radii. We thank Meredith Hughes for the helpful discussion on the stellar mass dependence of the planet-metallicity correlation. This research has made use of the NASA Exoplanet Archive, which is operated by the California Institute of Technology, under contract with the National Aeronautics and Space Administration under the Exoplanet Exploration Program.

\bibliography{mybib_JW_DF_PH5}

\clearpage

%
%
%


\clearpage


\newpage

\begin{landscape}

\begin{table}

\small
\caption{Comparison of the fractions of stars with detected planets for two metallicity groups in different planet radius range.   
\label{tab:planet_metallicity}}
\begin{tabular}{|p{3.5cm}|p{2.0cm}|p{2.0cm}|p{2.5cm}|p{2.0cm}|p{2.0cm}|p{2.8cm}|p{2.4cm}|}
\hline 
\multirow{2}{*}{Planet Radius} & \multicolumn{3}{|c|}{Metal-poor} & \multicolumn{3}{|c|}{Metal-rich} & \multirow{2}{*}{$\frac{(N_P/N_S)_{rich}}{(N_P/N_S)_{poor}}$} \\
\cline{2-7}
& $N_P$ & $N_S$ & $N_P/N_S$ & $N_P$ & $N_S$ & $N_P/N_S$ & \\

 \hline
\multirow{1}{*}{$R_P\leq1.7R_E$} & 51.0$\pm$8.1 & 53289$\pm$287 & 9.6$\pm$1.5 $\times$$10^{-4}$ & 82.0$\pm$10.2 & 49366$\pm$284 &  16.6$\pm$2.1 $\times$$10^{-4}$ & $1.72^{+0.19}_{-0.17}$ \\

\hline
\multirow{1}{*}{$1.7R_E\leq R_P\leq3.9R_E$} &  36.0$\pm$7.1 & 53289$\pm$287 & 6.8$\pm$1.3 $\times$$10^{-4}$ & 68.0$\pm$9.5 & 49366$\pm$284 &  13.8$\pm$1.9 $\times$$10^{-4}$ & $2.03^{+0.29}_{-0.26}$ \\

\hline
\multirow{1}{*}{$3.9R_E\leq R_P\leq22.0R_E$} & 3.0$\pm$2.1 & 53287$\pm$287 & 5.6$\pm$4.0 $\times$$10^{-5}$ & 24.0$\pm$5.4 & 49366$\pm$284 &  48.6$\pm$10.9 $\times$$10^{-5}$ & $9.30^{+5.62}_{-3.04}$ \\
\hline

\end{tabular}

\end{table}

\end{landscape}

\newpage

\end{document}